\documentstyle[12pt]{article}
\begin{document}
\baselineskip=16pt

\begin{center}
{\large STRUCTURAL INSTABILITY IN POLYACENE: A PROJECTOR QUANTUM MONTE CARLO
 STUDY }\\
\vspace{1cm}
{Bhargavi Srinivasan$^{2}$ and S. Ramasesha$^{2,3}$ }{\footnote{Electronic
address: bhargavi@sscu.iisc.ernet.in,~ramasesh@sscu.iisc.ernet.in\\}}\\
\vspace{0.5cm}

{\it
$^{2}$Solid State and Structural Chemistry Unit  \\
Indian Institute of Science, Bangalore 560 012, India \\
\vspace{0.5cm}
$^{3}$Jawaharlal Nehru Centre for Advanced Scientific Research \\
Jakkur Campus, Bangalore 560 064, India \\
}
\end{center}
\vspace{1cm}

\pagebreak
\clearpage

\begin{center}
{\bf{Abstract}}
\end{center}

We have studied polyacene within the Hubbard model to explore the 
effect of electron correlations on the Peierls' instability in a 
system marginally away from one-dimension. We employ the projector 
quantum Monte Carlo method to obtain ground state estimates of the 
energy and various correlation functions. We find strong
similarities between polyacene and polyacetylene
which can be rationalized from the real-space valence-bond
arguments of Mazumdar and Dixit. Electron correlations tend to enhance
the Peierls' instability in polyacene. This enhancement appears to 
attain a maximum at $U/t \sim 3.0$ and the maximum shifts to larger 
values when the alternation parameter is increased. The system shows 
no tendency to destroy the imposed bond-alternation pattern, as 
evidenced by the bond-bond correlations. The cis- distortion is seen 
to be favoured over the trans- distortion. The spin-spin
correlations show that undistorted polyacene is susceptible to
a SDW distortion for large interaction strength. The charge-charge 
correlations indicate the absence of a CDW distortion for the 
parameters studied. 

\section{Introduction}

The structural instabilities of one-dimensional systems, first predicted
by Peierls\cite{peierls} have been subjects of long standing interest 
to theoreticians and experimentalists.
Initial studies of the Peierls' instability focussed on
quasi-one-dimensional inorganic and organic charge-transfer  solids
such as KCP and TTF-TCNQ\cite{ctcom1,ctcom2,ctcom3}. However, later studies 
have concentrated 
on polyacetylene, the archetypal one-dimensional system.  Subsequent to the
initial H\"uckel theoretic descriptions of polyacetylene\cite{slh1}, 
much interest has been generated by the possibility of novel
solitonic and polaronic excitations in polyacetylene, the former suggested 
first in
the work of Pople and Wamsley\cite{pople1} and taken up again by 
Rice\cite{rice} and 
Su, Shrieffer and Heeger\cite{ssh}. The importance of including 
electron-electron
interactions in interpreting the electronic spectra of finite polyenes has been
demonstrated. Therefore there has been much interest in studying the
effect of electron correlations on the Peierls' instability in polyacetylene. 
Early work in this direction which described
dimerization within a real space valence-bond approach was carried out by
Coulson and Dixon\cite{coul1}.

The ordering of excited states has been studied in interacting models and the
correct ordering of excited states in long polyenes($2^1A_g$ below the 
optically allowed
$1^1B_u$ state) is obtained only at moderate to strong correlation 
strengths\cite{agbu}. The effect of electron correlations on the ground state
of polyacetylene is not obvious and contradictory results  were
obtained from different approximate theoretical approches. Mean-field and
Hartree-Fock approaches in general predict a decrease in dimerization
on including electron correlations even at the minimal level i.e. the
Peierls-Hubbard model. However, going beyond mean-field theory, 
a real-space picture of dimerization was gained by the valence-bond
analysis of Mazumdar and Dixit\cite{sumit}. They found that 
while correlations enhance dimerization and this persists over a large range
of correlation strength, the stabilization has a maximum for $U\sim 4t$.
Hirsch\cite{hirpeie} studied finite rings within the Peierls-Hubbard model
using a checkerboard Monte Carlo technique and obtained a similar maximum in the
enhancement.  The variational calculation of Baeriswyl and
Maki\cite{bae} agrees well with the quantum Monte Carlo results. The numerical 
renormalization group studies of Hayden and Mele\cite{hay} predict a
 similar maximum. 
Thus, it is now widely held that electron correlations enhance the 
instability in Peierls-Hubbard systems. The $XY$ spin model maps on to
a non-interacting spinless fermion model and is hence expected to show
a similar instability. The spin analogue of the Hubbard model is the
$XYZ$ model or Heisenberg model.  In spin-Peierls' systems, 
the effect of dimerization is to introduce a gap in the excitation 
spectrum\cite{bray}. The spin-Peierls' instability of the spin-half system 
has been extensively studied by probing  this gap. 

Another aspect of interest in Peierls' systems
is the effect of dimensionality on the instability. In non-interacting systems, 
this aspect has been studied quite generally in the framework of 
energy band theory and it is well recognized that the strength of the
instability depends upon the extent of nesting of the Fermi surface.  
The effect of increase in dimensionality on the spin-Peierls' 
instability has been explored by studying spin ladders and dimerized 
spin chains with next-nearest-neighbour interactions\cite{spinpeie}.

The effect of dimensionality on the Peierls-Hubbard system can also be 
studied by dealing with coupled Hubbard chains\cite{hublad}. However, 
Hubbard ladders 
have been studied extensively mainly to explore other kinds of 
instabilities such as the pairing instability.
Experimentally realizable systems which closely resemble Hubbard 
ladders are the polyacenes (Fig. (1a)). In the laboratory, polyacenes 
with up to seven rings have already been synthesized \cite{clar}. 
Early studies of the Peierls' instability in polyacenes were carried out 
by Salem and Longuet-Higgins\cite{slh2}, within the H\"uckel 
approximation. They considered
the stabilities of the cis- form of bond-alternated polyacene
(Fig. (1b)) and observed 
that the instability in these systems is only conditional, that is it 
could occur only below a critical force constant for a given electron-lattice
coupling strength. This is unlike the prediction in one-dimension 
where the distorted state is more stable than
the undistorted state independent of the magnitude of the force constant
and the strength of electron-lattice interaction. Boon\cite{boon} considered 
the trans- form of bond-alternated polyacene(Fig. (1c)) and argued that this
should be the more stable distortion. Misurkin and Ovchinnikov\cite{misov} 
predicted
that very long polyacenes should have an antiferromagnetic spin structure. 
The calculations of Whangbo, Woodward and Hoffmann\cite{whangbo} indicate 
that the trans- 
form is energetically stabilized. The more detailed CNDO calculations
of Tanaka et al\cite{tanaka} indicate that polyacene prefers the trans-
distorted ground state. 

There have been some theoretical studies on the electronically 
driven structural instability in polyacenes which include electron
correlations in the mean field approximation.  
Kivelson and Chapman\cite{KC} studied the possibility of bond alternation,
magnetic ordering and a superconducting transition as possible
broken symmetry states of polyacene.  Their mean-field calculations 
indicated that the bond-alternated state is not favoured. 
It has been conjectured that polyacene might show interesting conducting
properties in view of the small excitation gap. 
Electron 
correlations were explicitly included by O'Connor and Watts-Tobin\cite{ocwt} in
their study of polyacene which employed a modified Gutzwiller variational
ansatz. Their study showed that mean field phase diagram for the ground state
is only quantitatively modified and that the instability is only
conditional. However, the variational ansatz used by them makes assumptions
about the structure of the wavefunction and does not provide unbiased
correlations. Besides these studies there also exist other theoretical 
studies of polyacene which pertain to the origin of the band gap and
the crossover of these systems to the metallic state\cite{baldo,yamabe,bakshi}.

The issue of the effect of electron correlations on the instability 
predicted by non-interacting theories is far from being resloved. A proper
study of this aspect would require the use of reliable numerical techniques
for the ground state of the interacting model Hamiltonian for fairly
large system sizes.
 In recent years, the projector quantum Monte Carlo method has emerged
as a technique which is particularly well suited to study the Hubbard
model in higher dimensions.
In this paper we report our PQMC studies on the role of electron correlations 
on the different dimerization instabilities in polyacene. The PQMC method
provides a treatment of the Hubbard model that is exact within statistical
errors and makes no assumptions about the structure of the wavefunction
in estimating correlation functions. The technique also allows us to study
significantly larger system sizes than have been accessible through other
approches. The paper is organized as follows. We begin with a brief 
review of the results from the non-interacting model and then
present our results and discussions. We end with a summary of
our results in the last section.

\section{Results and Discussion}

Before discussing the interacting model, it would be worthwhile to 
present a brief outline of the analysis of the non-interacting model, 
due to Salem and Longuet-Higgins\cite{slh2}.
In the non-interacting picture, the Hamiltonian for polyacene (Fig. (1))
is given by
\begin{equation}
 \hat{H}_{0} =  \sum_{<ij>}\sum_{\sigma}t_{ij}
(c_{i\sigma}^{\dagger}c_{j\sigma}+h.c.)
\end{equation}
\noindent    
where $c^{\dagger}_{i\sigma}$ ($c_{i\sigma}$) is the
creation (annihilation) operator for an
electron with spin ${\sigma}$ in the Wannier orbital at the $i^{th}$ 
site and the summation $<ij>$ runs over bonded atom pairs. 
The bands of polyacene can be classified as symmetric or antisymmetric 
based on the
symmetry property (reflection about the plane bisecting the rungs)
of the MOs from which they are constructed. The ordering 
of the bands is such that the Fermi level lies between the antisymmetric 
and the symmetric bands. In the absence of any distortion,
for all values of the interchain coupling, there 
is a degeneracy, at the Fermi level between the top of the antisymmetric 
band and the bottom 
of the symmetric band. A symmetric 
distortion of the polyacene leads to a symmetric perturbing potential.
The matrix element of the perturbation between the degenerate states
at the Fermi level vanishes by symmetry. Therefore, the response of 
the electronic system to a finite perturbation does not diverge, in second
order. This divergence is essential for an unconditional Peierls'
distortion, as is found in the polyenes.
Thus, H\"uckel model studies indicate that the Peierls' instability
in polyacene is conditional  (the occurence of the distortion 
depends on the magnitude of the force constant).
Essentially the same result holds for a distortion in which the dimerization
of the top chain is out of phase with the dimerization of the bottom
chain (Fig. (1c)). A polymer with this type of distortion retains a $C_2$ symmetry
with the axis of symmetry being perpendicular to the molecular plane and
passing through the center of the polymer and the above reasoning carries 
through because the matrix element between states of different symmetry 
vanishes when the operator has the same symmetry as one of the states.

We now present results from projector quantum Monte Carlo
(PQMC) calculations of the ground states of polyacenes studied within
the framework of the Hubbard model.
\pagebreak
\clearpage
 The Hubbard Hamiltonian is given 
by,
\begin{eqnarray}
\hat{H} &=& \hat{H}_{0} + \hat{H}_{1} \\
 \hat{H}_{1} &=& U\sum_{i}\hat{n}_{i \uparrow}\hat{n}_{i \downarrow}\
\end{eqnarray}
\noindent          
where $\hat{H}_{0}$ is the H\"uckel Hamiltonian described above and
the notation is otherwise standard.
The PQMC method\cite{sorella} is a reliable method for obtaining
ground state estimates of various properties of the Hubbard model. 
In a PQMC calculation, a Trotter decomposition of the projection operator
$exp(-\beta \hat{H})$ is followed by a  discrete Hubbard-Stratonovich 
transformation which decouples the Hubbard interaction into
the interaction of fermions with Ising-like fields. Expectation values
are obtained as Monte Carlo averages by importance sampling the
Ising configurations. Estimates obtained from the PQMC method are
subject to a Trotter error and a statistical error, the first arising
from the Trotter decomposition and the second from the Monte Carlo
procedure. However, these can be controlled and it is possible to 
obtain accurate estimates of the energy and other correlation functions
from the PQMC method. Further details of the method can be obtained from
the literature\cite{sorella,imada,xtd}
However, a caveat in 
using the PQMC method is that a single-configurational trial wavefunction
is inadequate for 
"open-shell" systems\cite{bormann}, i.e. systems which have degenerate 
non-interacting 
ground states. Since the neutral polyacenes are "closed-shell", i.e. 
they posess unique non-interacting ground states, the PQMC method is 
expected to provide accurate estimates and would allow the simulation of
fairly large systems.

We study polyacenes with up to eleven rings, with periodic boundary 
conditions. We have considered the undistorted and cis- and trans- 
distorted (about the cross-links) forms shown in Fig. (1). We study 
the effect of bond-alternation
by imposing an alternation in the transfer integrals, $\delta$, as shown 
in Fig. (1). With this notation, the double lines between sites $i$ and
$j$ in Fig. (1) would correspond to shorter bonds between these sites. 
Thus hopping between sites $i$ and $j$
would have a transfer integral $(1+\delta)t_{ij}$. The transfer integrals 
corresponding to the other bonds are so modified as to keep the sum
of transfer integrals constant to allow reasonable comparison of energies. 
We have studied the effect of increasing correlation
strength, from weak through intermediate values ($U/t$=4.0) 
for alternation parameter values $\delta$= 0.05, 0.1 and 0.2. 

We have studied the instability in polyacene within the static lattice 
approximation. Our 
approach is to impose a bond alternation pattern and then to study the effect
of electron correlations. The strength of the Hubbard interaction is 
systematically increased from the H\"uckel limit and the results obtained
compared with the non-interacting results. These studies are carried out
for different values of the alternation parameter and for different types of
distortions. The issues we address are the following:
(i) the effect of electron correlations on the stability of the distorted 
states (ii) differences between systems with odd and even numbers of rings,
if any (iii) whether any qualitative change occurs on increasing the 
alternation parameter (iv) effect of interactions on the imposed bond 
order wave (BOW) (v) comparison with polyacetylene and (vi) interpretation 
in terms of real-space pictures along the lines of Mazumdar and Dixit for
polyacetylene rings.

We first use a n\"aive approach to the Peierls' distortion, entirely in 
terms of energetics. We study the gap
$ \Delta E_{A}(N,\delta,U)$ $\Bigl[ = E_{A}(N,\delta,U) - 
\Delta E(N,0,U)$; $A$= cis-, trans- $\Bigr]$
between the distorted
and undistorted state, varying the number of rings ($N$), the alternation
parameter ($\delta$) and the strength of electron correlations ($U$). In
Figs. (2) and (3), we present the variation of $\Delta E_{cis}(N,\delta,U)/N$ 
and $\Delta E_{trans}(N,\delta,U)/N$  with number of rings in the system.
With our definition, a negative value of $\Delta E$ would imply stabilization
of the distorted state. We observe that the distorted state is stabilized
with respect to the undistorted state even for $U/t=0.0$. At $U/t=0.0$,
increasing the alternation parameter $\delta$ tends to increase the
stabilization. At the level of the H\"uckel model, the cis- and trans-
distortions are  stabilized to the same extent. There is a difference
in the behaviour of systems with odd and even numbers of rings even within
the H\"uckel picture. Introducing electron-electron interactions appears
to increase the stabilization of the distorted state, for both cis- and
trans- distortions. We digress at this point to establish a connection
between polyacene and coupled polyacetylene chains, which will be useful
in interpreting most of the results.

We note that an $N$-ring polycene with periodic boundary conditions has
$4N$ sites. If we view an $N$-ring polyacene as coupled
polyacetylene chains with missing alternate rungs, it is evident that 
systems with odd numbers of rings would correspond to coupled $4n+2$
rings and that systems with even numbers of rings would correspond to
coupled $4n$ "polyacetylene" rings ($2n+1$ = $N$).  
The larger stabilization of the even-$N$ polyacenes relative to the odd-$N$
polyacenes (Figs. 2, 3 and 4) can be rationalized using the real space
arguments of Mazumdar and Dixit (MD). They showed that for cyclic polyenes,
in the space of covalent functions, the $4n$ ring systems have a 
stronger propensity for dimerization since the fraction of covalent
VB states that favour uniform bonds is zero for all $n$. However, for 
the $4n+2$ ring systems, this fraction decreases to zero
from a finite value as $n \rightarrow \infty$. This explains the opposite
trend in the stability of the even-$N$ polyacenes compared to the 
odd-$N$ polyacenes as a function of the system size, N upon introducing
dimerization. The data shown in Figs. (2 - 4)  clearly indicate
that marked differences exist between systems with even and odd numbers 
of rings. If we interpret a larger stabilization energy as being indicative
of a greater susceptibility to distortion, systems with even numbers
of rings are seen to be more susceptible to distortion and their
stabilization decreases with increasing system size. The
opposite trend can be noted for systems with odd numbers of rings from
Figs. (2 - 4). In polyacene, the most favourable extended conjugation 
pattern can be obtained by treating it as coupled polyacetylene
chains within the given structural framework. Introducing "double bond"
paths along the rungs would only lead to break in conjugation
and thus an overall destabilization. This would account for the
similarity in the results for polyacene and polyacetylene.

We now examine the effect of correlations on the stabilization of
polyacene systems with odd numbers of rings.  As can be seen from
Figs. (2) and (3), these systems are further stabilized by 
a non-zero $U/t$.  This stabilization is seen to increase initially with 
increasing $U/t$, but appears to reach a maximum at $U/t=3.0$, whereupon
the effect of $U/t$ is to relatively destabilize the distorted state. 
Furthermore, upon increasing $\delta$ to 0.2, it can be seen from Fig. (4) 
that the maximum value of the stabilization is not attained 
even at $U/t=4.0$. As shown by Mazumdar
and Dixit, introducing the Hubbard interaction increases
the tunneling barrier between the structures with two 
opposite phases for dimerization. This leads to an increase
in the energy difference between the dimerized and undimerized
structure upon increasing the Hubbard parameter $U/t$. This increased 
stabilization gradually reduces at larger on-site correlation
strengths as the energy scale in the problem changes from
the transfer integral $t$ to $J=2t^2/U$. 
Since the quantities presented here are differences in energies, they
are more difficult to measure accurately than just ground state energies.
Furthermore, it is well known that the PQMC  method becomes increasingly
inaccurate with increasing correlation strength. Since the system
shows the interesting turn-around behaviour in $U/t$ even at $\delta=0.1$,
we concentrate on this case without any loss in generality. 
We observe that systems with even numbers of rings also show an increase in
stabilization energy with increasing $U/t$. However, there is no discernible
maximum in the stabilization energy, for the system sizes and correlation 
strengths that we have studied. It is evident that the system size variations 
are strong even at 10 rings for polyacenes with even numbers of rings. 
However, the extrapolated stabilization energies for the infinite system size
obtained from the results for the even-N polyacenes do show a maximum 
between $U/t$= 3.0 and 4.0. However, it is not clear at what system size 
this effect will be observed for finite systems. 
The trends obeserved in the stabilization energy with $\delta$ 
and $U/t$ can be compared to the behaviour observed in polyacetylene. 
As mentioned previously, a variety of analyses, based on QMC methods, 
the real-space VB analysis, the variational approaches and numerical 
RG methods have indicated that the effect of electron correlations is 
to enhance the Peierls' instability in polyacetylene. 

A physical picture of the ground state can be obtained by analyzing 
the various correlation functions. The rest of our analysis 
concentrates on systems with odd numbers of rings and specifically on 
the system of 11 rings, the largest that we have studied. The 
correlation functions of smaller systems with odd and even numbers of 
rings also show similar behaviour and we believe that the results
of the chosen system are indeed representative of the infinite system.
We have computed the bond-bond, spin-spin and charge-charge
correlations of these systems. To study the existence of a BOW in 
systems which have degenerate ground states, the bond-bond correlation 
function defined as,
\begin{equation}
\langle b_{i} b_{j} \rangle = \langle \sum\limits_{\sigma} 
\bigl( a^{\dagger}_{i,\sigma}a_{i+1,\sigma} + h. c. \bigr)
\bigl( a^{\dagger}_{j,\sigma}a_{j+1,\sigma} + h. c. \bigr)\rangle,
\end{equation}
\noindent
should be studied. The bond order {\em per se} does not give 
information about the susceptibility to distortion 
in systems with possible degenerate distorted states, 
in the absence of an imposed distortion. Since the bond-bond
correlation describes relative distortions of bonds in the
system, in principle, its fourier transform would 
give the amplitudes for various kinds of BOW distortions.
The system has five bonds per unit cell and hence there
are fifteen possible bond-bond correlation functions. We
do not consider all these fifteen correlation functions
but instead consider only those correlation functions
that correspond to the distortions shown in Figs. (1b) and (1c).
For the numbering scheme shown in Fig. (1a), the bond which 
connects sites $i$ and $i+1$ is labelled $i$, with the exception
of those labelled $2N$ and $4N$, which connect sites
1 with $2N$ and $4N$ with $2N+1$. 

In Figs. (5a-c) we present the bond-bond correlations $\langle  b_{1}b_{j} 
\rangle$ for $U/t = 2.0$ for the undistorted and $\delta =0.1$ cis- and
trans- distorted forms of polyacene with 11 rings. To compare the effect
of increased correlation strength, we present these correlations for 
$U/t=4.0$, in Figs. (6a-c).  In these figures, $j=2,\ldots 2N$ label the 
bond-bond
correlation of the bond "1" with bonds on the upper chain and $j=2N+1 \ldots
4N$ label the correlation of bond "1" with bonds on the lower chain. From
Figs. (5a) and (6a) we see no evidence for bond-alternation in the
ground state of polyacene when it is not imposed in the Hamiltonian,
independent of the strength of correlations.
However in Fig. (5b) and (5c), we see that the bond-bond correlation
reflects the cis- and the trans- bond alternation imposed on the
system respectively.  Increasing $U/t$ to 4.0 does not bring about any
qualitative change in the picture (Figs. (6b-c)). However, the amplitude
for the cis- distortion is larger than that for the trans- distortion
for the same imposed bond alternation $\delta$. It is also seen that
on going from $U/t=2.0$ to $U/t=4.0$, the amplitude of the cis- distortion
increases slightly while that of the trans- distortion decreases. Although
energetically, the cis- and trans- distortions are favoured almost
equally, the bond-bond correlation functions indicate a larger 
susceptibility to distortion of the cis type. It would be interesting 
to study the bond-bond correlation function of the doped system, since 
any domain walls resulting from change over in the phase of the distortion 
would be very clearly indicated.  However, this would require the use of
a multi-configurational trial wavefunction to obtain accurate open-shell
ground state energies and would be the subject of a different study\cite{xtd}.

In Figs. (7a-c) we present the spin-spin correlations $4 \langle s_{1}^z
 s_{j}^z\rangle$, where $j=2,\ldots 4N$, for the numbering shown in 
Fig. (1a).  The correlation function falls off very rapidly for $U/t=2.0$.
However, we observe from Figs. (8a-c) that while the uniform system
starts developing antiferromagnetic fluctuations, the correlation length
appears extremely small for the bond-alternated states. It is interesting 
to note that the amplitude for spin density wave (SDW) distortion
is noticeable for the uniform case while in the distorted system
this amplitude is nearly vanishing. This indicates that in polyacenes
the BOW and SDW are mutually exclusive. The charge-charge correlations
shown in Figs. (9a-c) and (10a-c) show that the charge density fluctuation
is negligible and rules out charge density wave state in both the
distorted and the uniform polyacenes. 

\section{Summary}

We have studied polyacene systems with up to eleven rings within the 
Hubbard model, using the PQMC method. We compare our results with
those known for polyacetylene and find strong similarities. 
We find that systems with even and odd numbers of rings exhibit
different behaviour, as do polyacetylene systems with $4n$ and
$4n+2$ sites. We compare polyacene systems with odd
numbers of rings to coupled $4n+2$ polyacetylene rings
and systems with even numbers of rings to coupled $4n $ polyacetylene
rings. Electron correlations tend to enhance the effect of bond-alternation. 
This effect is seen to pass through a maximum for $U/t \sim 3.0$, the 
value for polyacetylene being  $U/t \sim 4.0$. Furthermore, the correlation 
strength at which this maximum occurs is seen to be shifted to larger 
values of $U/t$ with increase in the bond alternation parameter. A study 
of the bond-bond correlations indicates that the system has no tendency 
to destroy the imposed bond-alternation pattern.It also shows that the 
cis- distortion is favoured over the 
trans- distortion. The spin-spin
correlations show that the undistorted polyacene has a tendency to
form a SDW for large interaction strength. The charge-charge correlations 
show no evidence for any CDW distortion for the paramaters studied.

\noindent
{\bf{Acknowledgement}}: We acknowledge financial support from
the Department of Science and Technology, India and the IFCPAR,
under project 1308-4.
We thank Biswadeb Dutta of the JNCASR
for help with the computer systems.

\pagebreak
\clearpage
\begin{center}
{\bf{Figure Captions}}
\end{center}

\begin{enumerate}
\item{{\bf{Figure 1}}: Structure of (a) undistorted (b) cis- distorted
and (c) trans- distorted polyacene}.
\item{{\bf{Figure 2}}: (a) $\Delta E_{cis} (N,\delta,U)/N$ 
and (b) $\Delta E_{trans} (N,\delta,U)/N$, with $\delta$=0.05 for 
polyacenes with 3 to 11 rings.}
\item{{\bf{Figure 3}}: (a) $\Delta E_{cis} (N,\delta,U)/N$ 
and (b) $\Delta E_{trans} (N,\delta,U)/N$, with $\delta$=0.1 for 
polyacenes with 3 to 11 rings.}
\item{{\bf{Figure 4}}: (a) $\Delta E_{cis} (N,\delta,U)/N$ 
and (b) $\Delta E_{trans} (N,\delta,U)/N$, with $\delta$=0.2 for 
polyacenes with 3 to 11 rings.}
\item{{\bf{Figure 5}}: Bond-bond correlations vs. bond separation
for (a) undistorted (b)
cis- distorted ($\delta=0.1$) and 
(c) trans- distorted polyacene ($\delta=0.1$) for
$U/t$=2.0.}
\item{{\bf{Figure 6}}: Bond-bond correlations vs. bond separation 
for (a) undistorted (b)
cis- distorted ($\delta=0.1$) and
 (c) trans- distorted polyacene ($\delta=0.1$) for 
$U/t$=4.0.}
\item{{\bf{Figure 7}}: Spin-Spin correlations vs. intersite separation
for (a) undistorted (b)
cis- distorted ($\delta=0.1$) and (c) trans- distorted polyacene
($\delta=0.1$) for
$U/t$=2.0.}
\item{{\bf{Figure 8}}: Spin-Spin correlations vs. intersite separation
for (a) undistorted (b)
cis- distorted ($\delta=0.1$) and 
(c) trans- distorted polyacene ($\delta=0.1$) for
$U/t$=4.0.}
\item{{\bf{Figure 9}}: Charge-charge correlations vs. intersite
separation for (a) undistorted (b)
cis- distorted ($\delta=0.1$) and 
(c) trans- distorted polyacene ($\delta=0.1$) for
$U/t$=2.0.}
\item{{\bf{Figure 10}}: Charge-charge correlations vs. intersite
separation for (a) undistorted (b)
cis- distorted ($\delta=0.1$) and 
(c) trans- distorted polyacene ($\delta=0.1$) for 
$U/t$=4.0.}
\end{enumerate}

\begin{thebibliography}{} 
\bibitem{peierls} R. E. Peierls, {\em {Quantum Theory of Solids}}, Clarendon,
Oxford (1955).
\bibitem{ctcom1} A. Madhukar, Solid State Commun., {\bf{15}}, 921 (1974).
\bibitem{ctcom2} I. Egri, Solid State Commun., {\bf{22}}, 281 (1977).
\bibitem{ctcom3} I. I. Ukrainskii, Zh. Eksp. Teor. Fiz., {\bf{76}}, 760 (1979)
(Sov. Phys. JETP, {\bf{49}}, 381 (1979)).
\bibitem{slh1} H. C. Longuet-Higgins and L. Salem, Proc. Roy. Soc. A,
{\bf{251}}, 172 (1959).
\bibitem{pople1} J. A. Pople and S. H. Wamsley, Mol. Phys., {\bf{5}}, 15 (1962).
\bibitem{rice} M. J. Rice, Phys. Lett., {\bf{71A}}, 152 (1979); M. J. Rice and
J. Timonen, {\em{ibid.}}, {\bf{73A}}, 368 (1979).
\bibitem{ssh} W. P. Su, J. R. Shrieffer and A. J. Heeger, Phys. Rev. Lett.,
{\bf{42}} 1698 (1979); Phys. Rev. B {\bf{22}}, 2099 (1980), {\em{ibid.}}
{\bf{28}}, 1138E (1983).
\bibitem{coul1} C. A. Coulson and W. T. Dixon, Tetrahedron, {\bf{17}},
 215 (1961).
\bibitem{agbu} Z. G. Soos, S. Ramasesha and D. S. Galvao, Phys. Rev. Lett.,
{\bf{71}}, 1609 (1993).
\bibitem{sumit} S. Mazumdar and S. N. Dixit, Phys. Rev. Lett., {\bf{51}},
292 (1983); S. N. Dixit and S. Mazumdar, Phys. Rev. B {\bf{29}}, 1824 (1984).
\bibitem{hirpeie} J. E. Hirsch, Phys. Rev. Lett., {\bf{51}}, 296 (1983).
\bibitem{bae} D. Baeriswyl and K. Maki,  Phys. Rev. B {\bf {31}}, 6633 (1985).
\bibitem{hay} G. Hayden and E. Mele,  Phys. Rev. B {\bf {32}}, 6527 (1985).
\bibitem{bray} J. W. Bray, L. V. Interrante, I. S. Jacobs and J. C. Bonner,
in {\em { Extended Linear Chain Compounds}}, {\bf{3}}, Plenum, New York (1982),
p. 353.
\bibitem{spinpeie} H. J. Schulz, Phys. Rev. B {\bf{34}}, 6372 (1986);
I. Affleck, D. Gepner, H. J. Schulz and T. Ziman, J. Phys. A {\bf{22}}, 511 
(1989); D. Guo, T. Kennedy and S. Mazumdar, Phys. Rev. B {\bf{41}}, 9592 
(1990), S. K. Pati, S. Ramasesha and D. Sen, (unpublished).
\bibitem{hublad} R. M. Noack, S. R. White and D. J. Scalapino (unpublished),
preprint no. cond-mat/9601047.
\bibitem{clar} E. Clar, {\em{Polycyclic Hydrocarbons}}, Academic, New York
(1964).
\bibitem{slh2} L. Salem and H. C. Longuet-Higgins, Proc. Roy. Soc., A
 {\bf{255}}, 435 (1960).
\bibitem{boon} M. R. Boon, Theoret. Chim. Acta (Berl.), {\bf{23}}, 109 (1971).
\bibitem{misov} I. A. Misurkin and A. A. Ovchinnikov, Theoret. Chim. 
Acta (Berl.), {\bf{13}}, 115 (1969); Russ. Chem. Rev., {\bf{46}}, 967 (1971).
\bibitem{whangbo} M. -H. Whangbo, R. Hoffmann and R. B. Woodward,
Proc. Roy. Soc. A {\bf{366}}, 23 (1979).
\bibitem{tanaka} K. Tanaka, K. Ohzeki, S. Nankai, T. Yamabe and 
H. Shirakawa, J. Phys. Chem. Solids, {\bf{44}}, 1069 (1983).
\bibitem{KC} S. Kivelson and O. L. Chapman, Phys. Rev. B {\bf{28}},
7236 (1983).
\bibitem{ocwt} M. P. O'Connor and R. J. Watts-Tobin, J. Phys. C:Solid State
Phys., {\bf{21}}, 825 (1988).
\bibitem{baldo} M. Baldo, A. Grassi, R. Pucci and P. Tomasello,
J. Chem. Phys., {\bf{77}}, 2438 (1982);
M. Baldo, G. Piccito, R. Pucci and P. Tomasello, Phys. Lett., {\bf{95A}},
201 (1983).
\bibitem{yamabe}T. Yamabe, K. Tanaka, K. Ohzeki and S. Yata, Solid State
Commun., {\bf{44}}, 823 (1982).
\bibitem{bakshi} A. K. Bakshi and J. Ladik, Ind. J. Chem, {\bf{33A}},
494 (1994).
\bibitem{sorella} S. Sorella, E. Tosatti,  S. Baroni, R. Car and 
M. Parrinello, Int. J. Mod. Phys. B {\bf  1}, 993 (1988).
\bibitem{imada} M. Imada and Y. Hatsugai, J. Phys. Soc. Jpn.,
{\bf 58},  3752 (1989).
\bibitem{xtd} Bhargavi Srinivasan, S. Ramasesha and H. R. Krishnamurthy,
Phys. Rev. B {\bf{54}}, R2276 (1996); (unpublished), preprint no. cond-mat/
9609024.
\bibitem{bormann} D. Bormann, T. Schneider and M. Frick,
Z. Phys. B {\bf  87}, 1 (1992).
\end{thebibliography}
\end{document}